\documentclass[%
 aip,
 amsmath,amssymb,
 reprint,%
]{revtex4-1}

\usepackage{graphicx}
\usepackage{dcolumn}
\usepackage{bm}
\usepackage[utf8]{inputenc}
\usepackage[T1]{fontenc}
\usepackage{mathptmx}
\usepackage{etoolbox}
\usepackage{xcolor}

\makeatletter
\def\@email#1#2{%
 \endgroup
 \patchcmd{\titleblock@produce}
  {\frontmatter@RRAPformat}
  {\frontmatter@RRAPformat{\produce@RRAP{*#1\href{mailto:#2}{#2}}}\frontmatter@RRAPformat}
  {}{}
}%

\makeatother

\begin{document}

\preprint{AIP/123-QED}

\title[]{Load-dependent Hardness Prediction for Materials using Machine Learning}

\author{Madhubanti Mukherjee}
\affiliation{School of Materials Science and Engineering, Georgia Institute of Technology, Atlanta, Georgia 30332, United States}
\affiliation{Laboratory of Computational Chemistry and Biochemistry, Institute of Chemical Sciences and Engineering, Swiss Federal Institute of Technology (EPFL), Lausanne, Switzerland}
\author{Rampi Ramprasad}
\affiliation{School of Materials Science and Engineering, Georgia Institute of Technology, Atlanta, Georgia 30332, United States}
\author{Harikrishna Sahu$^*$}%
\affiliation{School of Materials Science and Engineering, Georgia Institute of Technology, Atlanta, Georgia 30332, United States}
\email{hsahu3@gatech.edu}

\begin{abstract}
Superhard materials are critical for wear-resistant and high-stress applications. Conventional approaches correlating hardness with elastic moduli derived from DFT calculations enable rapid screening but overlook the strong load dependence of hardness. In this work, machine learning (ML) models were developed using a large, curated dataset of load-dependent experimental Vickers hardness (H$_v$) measurements. Moderate correlation was observed between experimental and DFT-based H$_v$ values, whereas a single-task ML model trained solely on experimental data outperformed multi-task models that combined experimental and computed data. The superior performance of the single-task model highlights that explicit inclusion of indentation load, along with compositional, electronic, and structural descriptors, is essential and sufficient for accurate hardness prediction, beyond what can be achieved using DFT-accessible bulk and shear moduli alone (or in tandem with experimental data). These results emphasize the importance of high-quality experimental data and explicit inclusion of measurement conditions, particularly load, in the development of reliable hardness prediction models.
\end{abstract}

\maketitle

Hardness is a key mechanical property that underpins technologies ranging from cutting tools and abrasives to protective coatings and structural components.\cite{kaner2005designing,gao2005prediction} Superhard materials, typically defined by Vickers hardness (H$_v$) above 40 GPa,\cite{zhao2016recent,yeung2016ultraincompressible,kvashnin2019computational} are essential for extreme environments where resistance to plastic deformation is critical. The experimental discovery of such materials is labor-intensive, which motivates the development of predictive models. Semiempirical approaches that estimate H$_v$ from DFT-calculated elastic moduli enable high-throughput screening but rely on simplified assumptions that often fail under real conditions.\cite{jiang2011correlation,vsimuunek2006hardness,gao03:015502,tehrani2019hard} Machine learning (ML) models offer a flexible alternative with improved accuracy, yet most treat H$_v$ as a static property\cite{de2016statistical,isayev2017universal,mansouri2018machine,avery2019predicting,chen2021machine,muk24:10372}, neglecting the influence of applied load. 

While a few previous studies have incorporated load as an input parameter in machine learning models for hardness prediction and developed screening frameworks\cite{zha21:2005112}, the size and diversity of the dataset remain critical factors in assessing the applicability and reliability of such models. In order to ensure statistical robustness, it is important to utilize datasets that span a broad range of material compositions. Additionally, prediction-specific uncertainty quantification is particularly important for experimentally measured properties such as hardness, where variability arising from measurement conditions (such as load) can significantly influence reported values. In this context, probabilistic learning frameworks such as Gaussian Process Regression (GPR) provide reliable confidence estimation through principled uncertainty quantification. Another important objective is to understand whether commonly used DFT-based hardness proxies meaningfully improve model performance or introduce bias when combined with experimental data. Furthermore, empirical hardness models derived from elastic moduli are themselves subject to inherent physical limitations. In this context, a multitask learning framework is a useful approach to assess the relative contribution of these proxies and to identify which, if any, offer closer agreement with experimentally measured hardness or add predictive value to the model.

Here, we present a systematic evaluation of single-task versus multitask learning frameworks by explicitly including load as an input descriptor, allowing the model to capture its well-known dependence on hardness. Using a large experimental dataset spanning diverse crystal systems, compositions, and applied loads, we compared single-task (ST) and multi-task (MT) machine learning models built from purely experimental data and those incorporating DFT-based proxies to evaluate predictive performance. Notably, ST models trained solely on experimental data outperform all others, underscoring the limitations of DFT-based proxies in the capture of hardness physics.

\begin{figure*}
    \centering
    \includegraphics[width=1.0\linewidth]{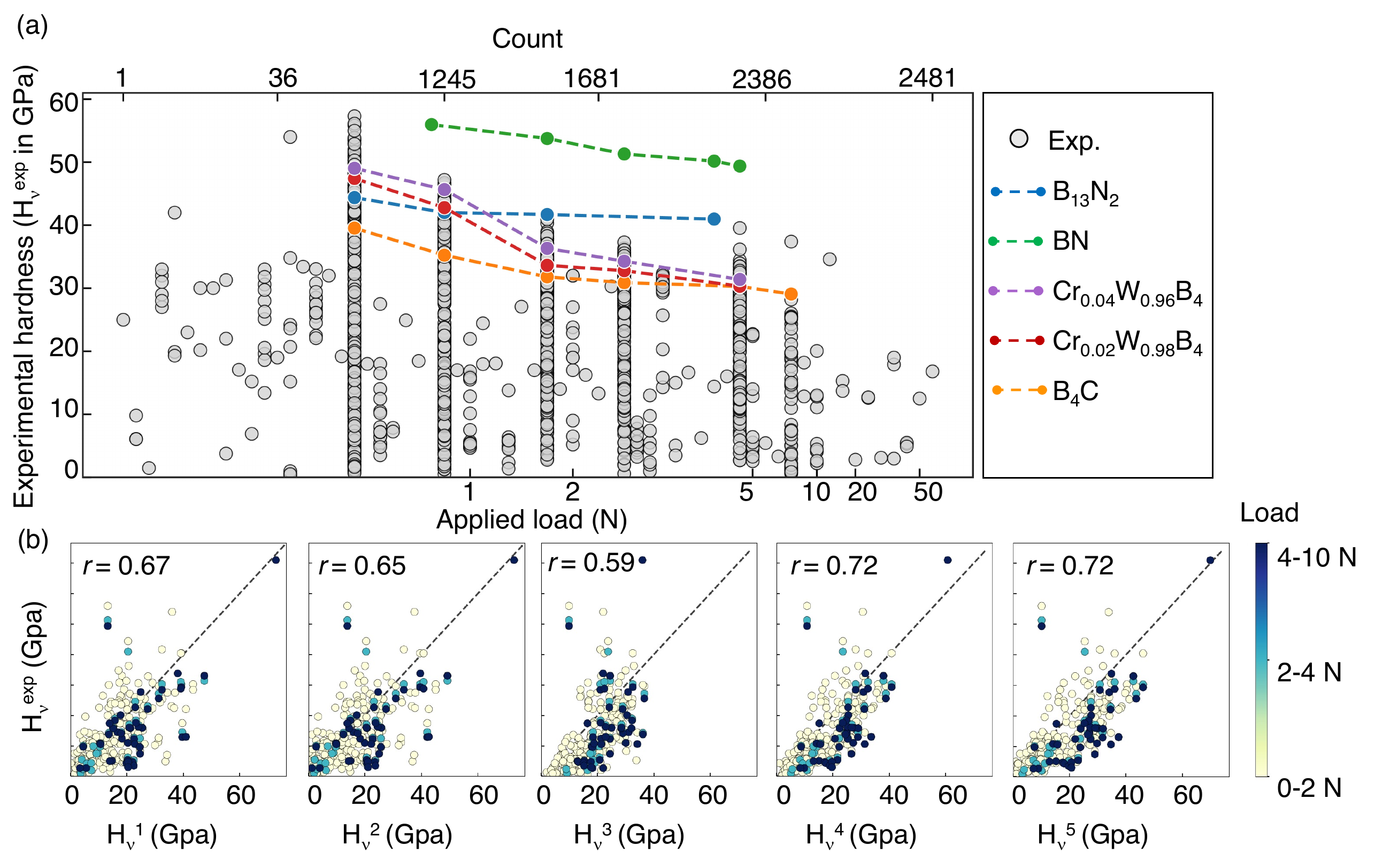}
    \caption{Hardness data showing (a) the distribution of experimental hardness values with respect to applied load and (b) the correlation between experimental and theoretically calculated hardness obtained using different approaches.}
    \label{fig2}
\end{figure*}

To develop a reliable, load-dependent hardness prediction framework, a comprehensive dataset of experimental H$_v$ measurements was curated, primarily from Refs. [\cite{zha21:2005112}, \cite{hic22:10003}, \cite{jaafreh2022machine}]. After removing duplicates with identical compositions and loads, the final dataset comprised 2,480 unique records covering 69 unique elements (the top 20 most frequent elements are shown in Figure S1(a)) with compositions containing up to six elements (ternary: 1209, binary: 742, quaternary: 428, 5+ elements: 97 and unary: 4 as shown in Figure S1(b)) and a total of 514 distinct chemical systems. The distribution of data across applied loads is depicted in Figure \ref{fig2}(a). As shown, the measurements extend beyond 50 N, with 0-2 N (1,646 points), 2-4 N (364), 4-10 N (444), 10-50 N (24), and $>$50 N (2), representing a realistic range of indentation conditions for model training and validation.

To establish a baseline for evaluating ML performance and integrate the computed data into MT learning, several widely used semiempirical models were examined that estimate H$_v$ from DFT-calculated elastic moduli. Specifically, five models proposed by Tian \cite{gao03:015502}, Chen \cite{che11:1275}, Jiang \cite{jia11:2287}, and Teter \cite{tet98:22} were considered:
\begin{equation}
H_{v}^{1} = 0.92(G/B)^{1.137}G^{0.708}
\end{equation}
\begin{equation}
H_{v}^{2} = 2(G^{3}/B^{2})^{0.585} - 3
\end{equation}
\begin{equation}
H_{v}^{3} = 0.0963B
\end{equation}
\begin{equation}
H_{v}^{4} = 0.1475G
\end{equation}
\begin{equation}
H_{v}^{5} = 0.1769G - 2.899
\end{equation}
These five equations, referred to as Models 1–5 throughout the manuscript, relate H$_v$ to the bulk modulus (B) and shear modulus (G), which represent resistance to volumetric and shear deformation, respectively. Using both in-house DFT calculations and data retrieved from the Materials Project\cite{jai13:011002,de2015charting,chen2021machine}, a dataset of 10,448 compositions was compiled, each assigned a nominal load of 0 N to reflect their theoretical nature.

Notably, 350 compositions overlap between the experimental and DFT-derived datasets, enabling direct comparison between measured and computed H$_v$ values for model benchmarking. Scatter plots comparing experimental and predicted H$_v$ obtained from the five semiempirical models, along with the corresponding Pearson correlation coefficients($r$), are shown in Figure \ref{fig2}(b). To ensure a meaningful comparison, the experimental data were restricted to measurements obtained under low indentation loads (typically $\leq$ 10 N), where plastic deformation more closely reflects the idealized conditions assumed in theoretical models. \cite{williams1956hooke,lan2014relationships,gorban2025determination}. The correlation between empirical predictions and experimental H$_v$ ranged from 0.59 to 0.72 across the models. This moderate agreement highlights two fundamental limitations: (i) the static formulation of these models, which neglects the strong load dependence of hardness, and (ii) their inherent simplifying assumptions of isotropy, defect-free structures, and the absence of microstructural effects. To further examine load effects, we analyzed the variation of experimental H$_v$ with applied load for representative materials (Figure \ref{fig2}(a)). The results reveal that H$_v$ can change by more than a factor of two with increasing load-behavior not captured by any empirical model, underscoring the need for predictive frameworks that explicitly learn and account for load-dependent effects.
\begin{figure}[hbt!]
    \centering
    \includegraphics[width=0.8\linewidth]{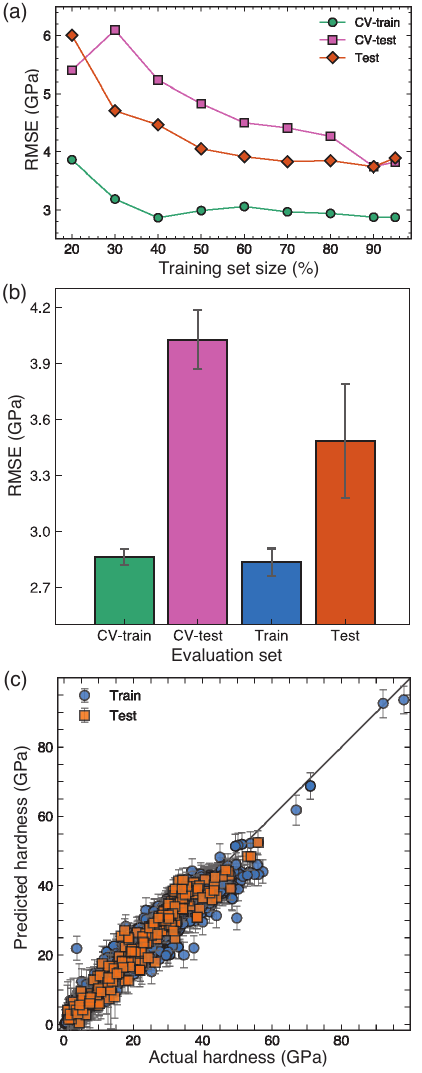}
    \caption{Performance of the single-task GPR model. (a) Learning curve showing RMSE (GPa) for cross-validation (CV)-train, CV-test, and independent test sets with increasing training-set size. (b) Average RMSEs with standard deviations over five random splits for CV-train, CV-test, train, and independent test sets. (c) Representative parity plot comparing predicted and actual hardness values for one split.}
    \label{fig3}
\end{figure}

To address the limitations of empirical approaches, a series of ML models was developed to establish a standardized modeling framework and evaluate the effect of dataset expansion and diversification on predictive performance. In total, six models were constructed: a ST model trained exclusively on experimental data, and five MT models that combine experimental data with computed hardness values obtained from the five semiempirical formulations given in Equations (1)–(5).

For feature generation, 60 compositional, electronic, and structural descriptors were computed based on the elemental composition of each compound. These descriptors include atomic number, mass, electronegativity, atomic radius, \textit{s}- and \textit{p}-orbital electron counts, and periodic table coordinates. Statistical measures such as mean, range, and weighted averages were applied to encode compositional complexity into the model space. In addition to these descriptors, the indentation load was explicitly incorporated as an input feature in all models to capture its critical influence on H$_v$. By embedding indentation load directly into the descriptor space, the model will be able to capture the transition from ideal elastic response at low or zero loads to deformation at higher loads, regimes that are generally averaged in elastic-modulus based empirical approaches. All ML models were developed using Gaussian Process Regression (GPR) training algorithm implemented in PolymRize™ platform\cite{PolymRize}, an advanced framework for rapid model development and evaluation. Each model was trained with 5-fold cross-validation (CV) and multiple random train–test splits to ensure statistical robustness, and performance was assessed using the root mean square error (RMSE) and coefficient of determination (R²).

Figure \ref{fig3} summarizes the performance of the developed GPR model using only the experimental dataset. The learning curves in Figure \ref{fig3}(a) show the average RMSE for the CV-train, CV-test, and independent test sets as a function of training set size. As expected, the RMSE values for both CV-test and test sets decrease with increasing training data, indicating improved generalization with larger datasets. Figure \ref{fig3}(b) presents the average RMSE from five different random splits, with error bars representing the standard deviation, while Figure S2 shows the corresponding R². The mean RMSE values for the CV-test and test sets are 4.03 and 3.49 MPa, respectively, with corresponding R² values of 0.91 and 0.93. A representative parity plot for the train and test sets is shown in Figure \ref{fig3}(c), and parity plots for all five random splits are provided in Figure S2.

The performance of the five MT GPR models, which combine experimental data with computed hardness values obtained from the five semiempirical approaches, is shown in Figure S3. As observed, the MT models incorporating computed data from the Chen and Tian formulations performed worse than the ST model, while the remaining three exhibited comparable RMSEs but lower R² values. These results indicate that incorporating empirically calculated H$_v$ values does not improve predictive performance for the materials considered. The absence of load dependence in the semiempirical H$_v$ values, along with their reliance on bulk and shear moduli, likely limits their physical relevance, yielding little to no improvement in MT learning performance. Several earlier ML models trained on hardness values derived from empirical correlations with DFT-calculated elastic moduli have exhibited consistently poorer performance relative to the ST model presented here.\cite{de2016statistical,isayev2017universal,avery2019predicting,chen2021machine,muk24:10372}. We note that the model is not expected to work well with extrapolation to loads or material chemistries outside the training domain, and the predictive performance may vary for materials with limited load coverage in the dataset or those exhibiting atypical or non-linear deformation behavior.

In this study, single-task and multi-task GPR models were developed to predict Vickers hardness (H$_v$) across a wide range of materials and indentation loads. The single-task model trained exclusively on experimental data achieved excellent predictive performance, with RMSE and R² values of 3.49 MPa and 0.93, respectively. In contrast, multi-task models that incorporated computed H$_v$ values from five semiempirical approaches alongside the experimental data showed no improvement and, in some cases, performed worse. These findings demonstrate that elastic moduli alone are insufficient for accurate hardness prediction, highlighting the importance of high-quality, load-dependent experimental data, along with compositional, electronic, and structural descriptors, for developing reliable and physically meaningful hardness prediction models.

\section{Supporting Information}
The Supplementary Material provides detailed performance analyses of the single-task GPR model trained solely on experimental data and the multitask models that integrate experimental and theoretical hardness data. The experimental hardness dataset, computational elastic moduli dataset, and hardness values calculated using five semiempirical models are available on GitHub at \url{https://github.com/Ramprasad-Group/polyVERSE}.

\section*{Conflicts of interest}
There are no conflicts to disclose.

\begin{acknowledgments}
The authors are grateful for the financial support of this work by the Office of Naval Research (grant number N00014-21-1-2258). The authors also thank ACCESS for computational support through allocation DMR080044.
\end{acknowledgments}

\bibliography{hardness}

@article{kaner2005designing,
  title={Designing superhard materials},
  author={Kaner, Richard B and Gilman, John J and Tolbert, Sarah H},
  journal={Science},
  volume={308},
  number={5726},
  pages={1268--1269},
  year={2005},
  publisher={American Association for the Advancement of Science}
}

@article{gao2005prediction,
  title={Prediction of new superhard boron-rich compounds},
  author={Gao, Faming and Qin, Xiujuan and Wang, Liqin and He, Yunhua and Sun, Guifang and Hou, Li and Wang, Wenyin},
  journal={The Journal of Physical Chemistry B},
  volume={109},
  number={31},
  pages={14892--14895},
  year={2005},
  publisher={ACS Publications}
}

@article{zhao2016recent,
  title={Recent Advances in Superhard Materials},
  author={Zhao, Zhisheng and Xu, Bo and Tian, Yongjun},
  journal={Annu. Rev. Mater. Res},
  volume={46},
  number={1},
  pages={383--406},
  year={2016}
}

@article{yeung2016ultraincompressible,
  title={Ultraincompressible, Superhard Materials},
  author={Yeung, Michael T and Mohammadi, Reza and Kaner, Richard B},
  journal={Annu. Rev. Mater. Res},
  volume={46},
  pages={465--485},
  year={2016},
  publisher={Annual Reviews}
}

@article{kvashnin2019computational,
  title={Computational Discovery of Hard and Superhard Materials},
  author={Kvashnin, Alexander G and Allahyari, Zahed and Oganov, Artem R},
  journal={J. Appl. Phys.},
  volume={126},
  number={4},
  pages={040901},
  year={2019},
  publisher={AIP Publishing LLC}
}

@article{jiang2011correlation,
  title={Correlation between hardness and elastic moduli of the covalent crystals},
  author={Jiang, Xue and Zhao, Jijun and Jiang, Xin},
  journal={Computational materials science},
  volume={50},
  number={7},
  pages={2287--2290},
  year={2011},
  publisher={Elsevier}
}

@article{vsimuunek2006hardness,
  title={Hardness of covalent and ionic crystals: first-principle calculations},
  author={{\v{S}}im{\r{u}}nek, Anton{\'\i}n and Vack{\'a}{\v{r}}, Ji{\v{r}}{\'\i}},
  journal={Phys. Rev. Lett.},
  volume={96},
  pages={085501},
  year={2006},
}

@article{gao03:015502,
  title = {Hardness of Covalent Crystals},
  author = {Gao, Faming and He, Julong and Wu, Erdong and Liu, Shimin and Yu, Dongli and Li, Dongchun and Zhang, Siyuan and Tian, Yongjun},
  journal = {Phys. Rev. Lett.},
  volume = {91},
  pages = {015502},
  year = {2003},
  doi = {10.1103/PhysRevLett.91.015502},
}

@article{tehrani2019hard,
  title={Hard and superhard materials: a computational perspective},
  author={Tehrani, Aria Mansouri and Brgoch, Jakoah},
  journal={Journal of Solid State Chemistry},
  volume={271},
  pages={47--58},
  year={2019},
  publisher={Elsevier}
}

@article{chen2021machine,
  title={Machine learning and evolutionary prediction of superhard BCN compounds},
  author={Chen, Wei-Chih and Schmidt, Joanna N and Yan, Da and Vohra, Yogesh K and Chen, Cheng-Chien},
  journal={npj Computational Materials},
  volume={7},
  number={1},
  pages={114},
  year={2021},
  publisher={Nature Publishing Group UK London}
}

@article{de2016statistical,
  title={A statistical learning framework for materials science: application to elastic moduli of k-nary inorganic polycrystalline compounds},
  author={De Jong, Maarten and Chen, Wei and Notestine, Randy and Persson, Kristin and Ceder, Gerbrand and Jain, Anubhav and Asta, Mark and Gamst, Anthony},
  journal={Scientific reports},
  volume={6},
  number={1},
  pages={34256},
  year={2016},
  publisher={Nature Publishing Group UK London}
}

@article{isayev2017universal,
  title={Universal fragment descriptors for predicting properties of inorganic crystals},
  author={Isayev, Olexandr and Oses, Corey and Toher, Cormac and Gossett, Eric and Curtarolo, Stefano and Tropsha, Alexander},
  journal={Nature communications},
  volume={8},
  number={1},
  pages={15679},
  year={2017},
  publisher={Nature Publishing Group UK London}
}

@article{mansouri2018machine,
  title={Machine learning directed search for ultraincompressible, superhard materials},
  author={Mansouri Tehrani, Aria and Oliynyk, Anton O and Parry, Marcus and Rizvi, Zeshan and Couper, Samantha and Lin, Feng and Miyagi, Lowell and Sparks, Taylor D and Brgoch, Jakoah},
  journal={Journal of the American Chemical Society},
  volume={140},
  number={31},
  pages={9844--9853},
  year={2018},
  publisher={ACS Publications}
}

@article{avery2019predicting,
  title={Predicting superhard materials via a machine learning informed evolutionary structure search},
  author={Avery, Patrick and Wang, Xiaoyu and Oses, Corey and Gossett, Eric and Proserpio, Davide M and Toher, Cormac and Curtarolo, Stefano and Zurek, Eva},
  journal={npj Computational Materials},
  volume={5},
  number={1},
  pages={89},
  year={2019},
  publisher={Nature Publishing Group UK London}
}

@article{jaafreh2022machine,
  title={Machine learning guided discovery of super-hard high entropy ceramics},
  author={Jaafreh, Russlan and Kang, Yoo Seong and Kim, Jung-Gu and Hamad, Kotiba},
  journal={Materials Letters},
  volume={306},
  pages={130899},
  year={2022},
  publisher={Elsevier}
}

@article{williams1956hooke,
  title={Hooke's Law and the Concept of the Elastic Limit},
  author={Williams, E},
  journal={Annals of Science},
  volume={12},
  number={1},
  pages={74--83},
  year={1956},
  publisher={Taylor \& Francis}
}

@article{lan2014relationships,
  title={On the relationships between hardness and the elastic and plastic properties of isotropic power-law hardening materials},
  author={Lan, Hongzhi and Venkatesh, TA},
  journal={Philosophical Magazine},
  volume={94},
  number={1},
  pages={35--55},
  year={2014},
  publisher={Taylor \& Francis}
}

@article{gorban2025determination,
  title={Determination of Material Hardness Characteristics at the Elastic Limit by Instrumented Indentation},
  author={Gorban, VF and Lugovyi, MI and Verbylo, DG},
  journal={Strength of Materials},
  volume={57},
  number={3},
  pages={567--572},
  year={2025},
  publisher={Springer}
}

@article{che11:1275,
title = {Modeling hardness of polycrystalline materials and bulk metallic glasses},
journal = {Intermetallics},
volume = {19},
pages = {1275-1281},
year = {2011},
doi = {https://doi.org/10.1016/j.intermet.2011.03.026},
author = {Xing-Qiu Chen and Haiyang Niu and Dianzhong Li and Yiyi Li},
}

@article{jia11:2287,
title = {Correlation between hardness and elastic moduli of the covalent crystals},
journal = {Computational Materials Science},
volume = {50},
number = {7},
pages = {2287-2290},
year = {2011},
doi = {https://doi.org/10.1016/j.commatsci.2011.01.043},
author = {Xue Jiang and Jijun Zhao and Xin Jiang},
}

@article{tet98:22,
title = {Computational Alchemy: The Search for New Superhard Materials},
journal = {MRS Bull.},
volume = {23},
pages = {22--27},
year = {1998},
author = {Teter, David M.},
}

@article{muk24:10372,
author = {Mukherjee, Madhubanti and Sahu, Harikrishna and Losego, Mark D. and Gutekunst, Will R. and Ramprasad, Rampi},
title = {Informatics-Driven Design of Superhard B–C–O Compounds},
journal = {ACS Appl. Mater. Interfaces},
volume = {16},
pages = {10372-10379},
year = {2024},
doi = {10.1021/acsami.3c18105},
}

@article{zha21:2005112,
author = {Zhang, Ziyan and Mansouri Tehrani, Aria and Oliynyk, Anton O. and Day, Blake and Brgoch, Jakoah},
title = {Finding the Next Superhard Material through Ensemble Learning},
journal = {Adv. Mater.},
volume = {33},
pages = {2005112},
doi = {https://doi.org/10.1002/adma.202005112},
year = {2021}
}

@article{hic22:10003,
author = {Hickey, Jacob C. and Brgoch, Jakoah},
title = {The Limits of Proxy-Guided Superhard Materials Screening},
journal = {Chem. Mater.},
volume = {34},
pages = {10003-10010},
year = {2022},
doi = {10.1021/acs.chemmater.2c02390},
}

@misc{PolymRize,
  author       = {{Matmerize, Inc.}},
  title        = {{PolymRize\texttrademark: An informatics platform for polymer property prediction and design using machine learning.}},
  howpublished = {\url{https://polymrize.matmerize.com/}},
  year         = {2025}
}

@article{jai13:011002,
    author = {Jain, Anubhav and Ong, Shyue Ping and Hautier, Geoffroy and Chen, Wei and Richards, William Davidson and Dacek, Stephen and Cholia, Shreyas and Gunter, Dan and Skinner, David and Ceder, Gerbrand and Persson, Kristin A.},
    title = {Commentary: The Materials Project: A materials genome approach to accelerating materials innovation},
    journal = {APL Materials},
    volume = {1},
    pages = {011002},
    year = {2013},
    doi = {10.1063/1.4812323},
}

@article{de2015charting,
  title={Charting the Complete Elastic Properties of Inorganic Crystalline Compounds},
  author={De Jong, Maarten and Chen, Wei and Angsten, Thomas and Jain, Anubhav and Notestine, Randy and Gamst, Anthony and Sluiter, Marcel and Krishna Ande, Chaitanya and Van Der Zwaag, Sybrand and Plata, Jose J and others},
  journal={Sci. Data},
  volume={2},
  number={1},
  pages={150009},
  year={2015},
  publisher={Nature Publishing Group}
}

\end{document}